\documentclass[multphys,vecphys]{svmult}

\usepackage{makeidx}     % allows index generation
\usepackage{graphicx}    % standard LaTeX graphics tool
\usepackage{multicol}    % used for the two-column index
\usepackage{url}         % to fix special characters in email
\makeindex             % used for the subject index
\begin{document}

\title*{From Molecular Cores to Planet-forming Disks with SIRTF }
% Use \titlerunning{Short Title} for an abbreviated version of
% your contribution title if the original one is too long
\author{Neal J. Evans II\inst{1}\and
The c2d Team\inst{2}}
% Use \authorrunning{Short Title} for an abbreviated version of
% your contribution title if the original one is too long
\institute{The University of Texas at Austin, Department of Astronomy,
       1 University Station C1400, Austin, Texas 78712--0259
\texttt{nje@astro.as.utexas.edu}
\and see the c2d website for the full list of team members 
\texttt{http://peggysue.as.utexas.edu/SIRTF/}}
%
% Use the package "url.sty" to avoid
% problems with special characters
% used in your e-mail or web address
%
\maketitle

%\section{Section Heading}
%\label{sec:1}
% Always give a unique label
% and use \ref{<label>} for cross-references
% and \cite{<label>} for bibliographic references
% use \sectionmark{}
% to alter or adjust the section heading in the running head
%Your text goes here. Use the \LaTeX\ automatism for your citations
%\cite{monograph}.

%
%
% BibTeX users please use
% \bibliographystyle{}
% \bibliography{}
%
% Non-BibTeX users please follow the syntax
% the syntax of "referenc.tex" for your own citations

% General purpose macros

\newcommand{\ee}[1]{\mbox{${} \times 10^{#1}$}}% scientific number format
\newcommand{\eten}[1]{\mbox{$10^{#1}$}}% power of ten

%macros for RA and Dec
%\newcommand{\h}{\mbox{{$^h$}}
%\newcommand{\m}{\mbox{{$^m$}}
%\newcommand{\s}{\mbox{{$^s$}}
\newcommand{\degree}{\mbox{$^{\circ}$}}
\newcommand{\arcmin}{\mbox{$^{\prime}$}}
\newcommand{\am}{\mbox{\arcmin}}
\newcommand{\as}{\mbox{\arcsec}}

%macros for distance, volume, speed
\newcommand{\kms}{\mbox{km s$^{-1}$}}% km/s
\newcommand\cmv{\mbox{cm$^{-3}$}}
\newcommand\cmc{\mbox{cm$^{-2}$}}
\newcommand\cmdv{\mbox{cm$^{-2}$ (\kms)$^{-1}$}}
\newcommand{\um}{$\mu$m}
\newcommand{\micron}{$\mu$m}

%macros for commonly used symbols
\newcommand{\x}{\mbox{${}\times{}$}}
\newcommand\tto{\mbox{$\rightarrow$}}
\newcommand\about{\mbox{$\sim$}}
%macros to avoid typing headache and needless acronyms simultaneously
\newcommand{\iras}{\mbox{\it IRAS}}
\newcommand{\iso}{\mbox{\it ISO}}
\newcommand{\sws}{\mbox{\rm SWS}}
\newcommand{\ISO}{\mbox{\it ISO}}
\newcommand{\ISOCAM}{\mbox{\rm ISOCAM}}
\newcommand{\sirtf}{\mbox{\it SIRTF}}
\newcommand{\SIM}{\mbox{\it SIM}}
\newcommand{\herschel}{\mbox{\it Herschel}}
\newcommand{\ngst}{\mbox{\it JWST}}
\newcommand{\sofia}{\mbox{\it SOFIA}}
\newcommand{\sma}{\mbox{SMA}}
\newcommand{\alma}{\mbox{ALMA}}
\newcommand{\carma}{\mbox{CARMA}}
\newcommand{\tpf}{\mbox{\it TPF/Darwin}}
\newcommand{\hst}{\mbox{\it HST}}
\newcommand{\rosat}{\mbox{\it ROSAT}}
\newcommand{\irac}{\mbox{\rm IRAC}}
\newcommand{\mips}{\mbox{\rm MIPS}}
\newcommand{\irs}{\mbox{\rm IRS}}
\newcommand{\ssc}{\mbox{\rm SSC}}
\newcommand{\gto}{\mbox{\rm GTO}}
\newcommand{\feps}{\mbox{\rm FEPS}}
\newcommand{\spot}{\mbox{\rm SPOT}}
\newcommand{\simba}{\mbox{\rm SIMBA}}
\newcommand{\sest}{\mbox{\rm SEST}}
\newcommand{\scuba}{\mbox{\rm SCUBA}}
\newcommand{\jcmt}{\mbox{\rm JCMT}}
\newcommand{\cso}{\mbox{\rm Caltech Submillimeter Observatory}}
\newcommand{\mambo}{\mbox{\rm MAMBO}}
\newcommand{\complete}{\mbox{\rm COMPLETE}}
\newcommand{\bolocam}{\mbox{\rm Bolocam}}
\newcommand{\iram}{\mbox{\rm IRAM}}
\newcommand{\mm}{millimeter}
\newcommand\submm{submillimeter}
\newcommand\smm{submillimeter}
\newcommand\fir{far-infrared}
\newcommand\mir{mid-infrared}
\newcommand\nir{near-infrared}
\newcommand\uv{ultraviolet}
\newcommand{\sfr }{\mbox{$\dot M_{\star}$}}
\newcommand\sed{spectral energy distribution}
\newcommand{\lsun}{\mbox{L$_\odot$}}% Lsun
\newcommand{\msun}{\mbox{M$_\odot$}}% Msun
\newcommand{\mmoon}{\mbox{M$_{Moon}$}}% Msun
\newcommand{\ta}{{$T_A^*$}}
\newcommand{\tex}{\mbox{$T_{\rm ex}$}}
\newcommand{\tmb}{\mbox{$T_{\rm mb}$}}
\newcommand{\tr}{\mbox{$T_R$}}
\newcommand{\tk}{\mbox{$T_K$}}
\newcommand{\td}{\mbox{$T_d$}}
\newcommand{\lbol}{\mbox{$L_{bol}$}} % bolometric luminosity
\newcommand{\tbol}{\mbox{$T_{bol}$}} % bolometric temperature
\newcommand{\dv}{\mbox{$\Delta v$}}
\newcommand{\n}{\mbox{$n$}}
\newcommand{\nbar}{\mbox{$\overline{n}$}}
\newcommand{\mv}{\mbox{$M_V$}} % virial mass
\newcommand{\mc}{\mbox{$M_N$}} % column density mass
\newcommand{\mn}{\mbox{$M_n$}} % density mass
\newcommand{\meanl}{\mbox{$\langle l \rangle$}} % mean size
\newcommand{\meandev}{\mbox{$\langle \delta \rangle$}} % mean deviation
\newcommand{\meanar}{\mbox{$\langle a/b \rangle$}} % mean aspect ratio
\newcommand{\mean}[1]{\mbox{$\langle#1\rangle$}} %generic mean for defined qu.
\newcommand{\opacity}{\mbox{$\kappa(\nu)$}} % opacity as func. of freq.
\newcommand{\av}{\mbox{$A_V$}} % Visual Extinction
\newcommand{\bperp}{\mbox{$B_{\perp}$}} % Projection of B on plane of sky
\newcommand{\rinf}{\mbox{$r_{inf}$}} % infall radius
\newcommand{\fsmm}{\mbox{$L_{smm}/L_{bol}$}} % submm lum over bol. luminosity
\newcommand{\lsmm}{\mbox{$L_{smm}$}} % luminosity longward of 350 mic.
\newcommand{\alphanir}{\mbox{$\alpha_{NIR}$}} % spectral index 2-20 mic.
\newcommand{\isrf}{\mbox{\rm{ISRF}}}

%macros for molecule names
\newcommand{\hh}{\mbox{{\rm H}$_2$}}
\newcommand{\form}{H$_2$CO}
\newcommand{\water}{H$_2$O}
\newcommand{\ammonia}{\mbox{{\rm NH}$_3$}}
\newcommand{\coo}{$^{13}$CO}
\newcommand{\cooo}{C$^{18}$O}
\newcommand{\coooo}{C$^{17}$O}
\newcommand{\hcop}{HCO$^+$}
\newcommand{\hcopi}{H$^{13}$CO$^+$}
\newcommand{\dcop}{DCO$^+$}
\newcommand{\nnhp}{N$_2$H$^+$}
\newcommand{\nthp}{N$_2$H$^+$}
\newcommand{\jj}[2]{\mbox{$J = #1\rightarrow#2$}}% J = i -> f

%Macros for this paper
\newcommand{\Inu}{\mbox{$I_{\nu}(b)$}}
\newcommand{\Snu}{\mbox{$S_{\nu}$}}
\newcommand{\Bnu}{\mbox{$B_{\nu}(\Td)$}}
\newcommand{\nInu}{\mbox{$I^{norm}(b)$}}
\newcommand{\nInum}{\mbox{$I^{norm}_{mod}(b)$}}
\newcommand{\kappanu}{\mbox{$\kappa_{\nu}$}}
\newcommand{\Td}{\mbox{$T_{d}$}}
\newcommand{\Tk}{\mbox{$T_{K}$}}
\newcommand{\Tdr}{\mbox{$T_{d}(r)$}}
\newcommand{\Tkr}{\mbox{$T_{K}(r)$}}
\newcommand{\rhor}{\mbox{$\rho (r)$}}
\newcommand{\ppc}{pre-protostellar core}
\newcommand{\m}{\mbox{$m$}}
\newcommand{\p}{\mbox{$p$}}
\newcommand{\nc}{\mbox{$n_c$}}
\newcommand{\chisq}{\mbox{$\chi_r^2$}}
\newcommand{\beam}{\mbox{$\theta_{mb}$}}
\newcommand{\router}{\mbox{$r_o$}}
\newcommand{\rinner}{\mbox{$r_i$}}
\newcommand{\rflat}{\mbox{$r_{flat}$}}
\newcommand{\mjup}{M$_{Jup}$}
\newcommand{\mearth}{M$_{Earth}$}
\newcommand{\pms}{pre-main-sequence}
\newcommand{\tnm}[1]{\mbox{\tablenotemark{#1}}}
\newcommand{\pasp}{Pub. Astr. Soc. Pac.}
\newcommand{\apj}{Astrophys. J.}
\newcommand{\apjs}{Astrophys. J. Suppl.}
\newcommand{\apjl}{Astrophys. J. (Letters)}

\begin{abstract}

The \sirtf\ mission and the Legacy programs will provide coherent
data bases for extra-galactic and Galactic science that will rapidly
become available to researchers through a public archive. The 
capabilities of \sirtf\ and the six legacy programs are described briefly. 
Then the cores to disks (c2d) program is described in more detail.
The c2d program will use all three \sirtf\
instruments (\irac, \mips, and \irs ) to observe sources
from molecular cores to protoplanetary
disks, with a wide range of cloud masses, stellar masses, and
star-forming environments. 
The \sirtf\ data will stimulate many
follow-up studies, both with \sirtf\ and with other instruments.

\end{abstract}

%\keywords{surveys: infrared --- stars: formation  --- 
%planetary systems: formation --- planetary systems: protoplanetary disks ---
%ISM: dust, extinction --- ISM: clouds }

\section{Why \sirtf ?}
\label{sec:1}

Much of the radiant energy in the Universe lies in the infrared region,
and infrared observations are well suited to the study of distant starburst 
galaxies and star formation, where dust controls the flow of energy.
Observational studies of galaxies and star formation have generally suffered
from one or more of the following problems: biased samples,
inadequate sensitivity, inadequate spatial resolution, or incomplete spectral 
data. 

The Space Infrared Telescope Facility (\sirtf )
\cite{Gallagher 2003} will provide greatly improved
capabilities in the infrared region.
\sirtf\ is the last of the series of four great observatories that began with
the Hubble Space Telescope and continued with the Compton Gamma Ray 
Observatory and the Chandra X-ray Observatory. \sirtf\ covers the wavelength
region from 3.6 to 160 \micron\ with background-limited performance. Its
85-cm diameter beryllium mirror is cooled below 5.5 K, and the cryogen lifetime
should be between 2.5 and 5 years. It was launched on August 25, 2003 into
an earth-trailing solar orbit.

The instrument complement includes two imaging array instruments and a
spectrometer. 
The {\it InfraRed Array Camera} or \irac, covers 3.6 to 8 \micron\ 
\cite{Fazio 1998} in four bands;
the {\it Multiband Imaging Photometer for SIRTF} or \mips, covers 24 to
160 \micron\ \cite{Englebracht 2000} in 3 bands;
and
the {\it InfraRed Spectrometer}, or \irs, supplies spectroscopy from 
5.3 to 40 \micron\ with resolving power $R = 60-120$ and from 
10 to 37 \micron\ with $R = 600$ 
\cite{Houck 2000}.
The field of view of the imagers is 5\am\ by 5\am\ except
at 160 \micron, where the field of view is 0.5\am\ by 5\am.
\mips\ also has an $R = 15$ spectrophotometric mode between
50 and 100 \micron.

\section{The Legacy Programs}
\label{sec:2}

Observing time on \sirtf\ is divided between Guaranteed Time Observers
(\gto s), Legacy programs, and General Observer (GO) programs. The legacy
program aims to ensure that coherent data bases of general and 
lasting value will be obtained during the limited mission lifetime. To enhance
opportunities for follow-up 
with \sirtf, the legacy data have no proprietary period; after the archive
has opened, about 8 months after launch, data will become available
to the public and the legacy teams simultaneously. The legacy teams will
provide improved data and, in many cases, ancillary data from other
wavelengths and tools for analysis. Legacy programs have typically 
hundreds of hours of observing time and, insofar as is possible, will
be observed early in the mission.

There are three extra-galactic legacy programs and three Galactic legacy
programs. The Great Observatories Origins Deep Survey (GOODS) will survey
300 square arcminutes with \irac\ and \mips\ (24 \micron\ only) overlapping
the Hubble and Chandra deep fields. New imaging with Hubble's Advanced
Camera for Surveys has been obtained.
Extreme starburst galaxies may be detected to $z = 6$ \cite{Dickinson 2003}.
The \sirtf\ Wide-area Infrared Extragalactic Survey (SWIRE) will map 100
square degrees at high Galactic latitudes with \irac\ and \mips, 
detecting dusty star-forming galaxies and AGNs to 
$z \sim 3$ \cite{Lonsdale 2003}.
The \sirtf\ Nearby Galaxies Survey (SINGS) will study 75 nearby galaxies
using all the \sirtf\ instruments, providing a template for understanding
more distant galaxies \cite{Kennicutt 2003}.

Turning to the Galactic legacy projects, the \sirtf\ Galactic Plane Survey
(GLIMPSE) will survey 240 square degrees of the inner Galactic plane with
\irac\ \cite{Benjamin 2003}. The c2d project (described in more detail
below) studies nearby low mass star formation and disk evolution 
\cite{Evans 2003}. The study of disk evolution is continued to later stages
by the Formation and Evolution of Planetary Systems (FEPS) team 
\cite{Meyer 2002}. The FEPS team will obtain images and spectroscopy of 300 
stars with ages from 3 Myr to 1 Gyr.

\section{The Cores to Disks Legacy Program}
\label{sec:3}

The Cores to Disks (c2d) team includes 11 co-investigators, 34 associates, 
and 20 affiliates.
Associates will work directly on \sirtf\ data; affiliates are adding
data at other wavelengths. The goal is to obtain as complete a data base
as possible for nearby ($d < 350$ pc) low-mass star formation. The data base
will be used to follow the evolution of molecular cloud material from starless
cores to stars surrounded by planet-forming disks. The c2d team will observe
stars with ages up to 3-10 Myr and will obtain spectroscopy of objects with 
a range of ages to trace the evolution in the nature of the dust and gas.

Using 400 hours and all three \sirtf\ instruments, the c2d 
program will provide the most complete data base of infrared observations
for low-mass star formation. Ancillary and complementary data will be
collected at wavelengths from X-ray to radio for these same regions.
We will map 5 large nearby molecular clouds and 156 compact
molecular cores with both \irac\ and \mips\ (275 hours). 
Photometry will be obtained for 190 stars with \irac\ and \mips\ (50 hours).
Spectroscopy will be obtained with \irs\ for at least 170 targets (75 hours).

%cores.tex  This covers small cores and large clouds, along with
%ancillary data.

\subsection{A Survey of Nearby Molecular Clouds and Cores} \label{cores}

The \sirtf\ mission offers an unprecedented opportunity to determine 
the stellar content of the nearest star-forming molecular clouds, the 
distributions of their youngest stars and substellar objects, and the 
properties of their circumstellar envelopes and disks.  
Our program will address the questions in the following paragraph.

In the large clouds, we will study the distribution of 
the youngest stars and substellar objects to understand the conditions
needed for their formation.
Do isolated cores with associated stars preferentially harbor 
single stars or small stellar groups?  
Can brown dwarfs form in the way that ordinary low-mass stars do, 
or are they ejected from small stellar groups? If very young brown
dwarfs are found throughout the cloud, surrounded by envelopes, an extension
of low-mass star formation processes into the substellar regime would be viable.
Deep \irac\ imaging of dense cores will reveal their detailed 
structure by mapping the extinction of background stars.  This 
structure will be compared with that derived from mapping emission from 
the dust continuum.
The incidence of circumstellar disks in complexes and in 
isolated cores will elucidate the role of the environment in disk formation.  
Our less biased sample will allow more robust measures of the 
lifetimes of various stages in the star formation process.
Rare objects and systems in rapid phases of evolution may be present in
a large sample.  We may, for example, discover the
short-lived phase in which the first, molecular core forms, an
as yet unobserved stage of star formation \cite{Boss 1995}, 
\cite{Masunaga 1998}.

%Describe the sample of clouds and cores, reasons for choosing them,
%systematics of distances, types of conditions, etc.

We chose a sample of clouds and cores that will sample a wide range
of conditions in order to
separate the effects of evolution from other variables.
We target five large
complexes known to be forming stars in isolation, in groups, and in
clusters, and 156 small, isolated cores -- 110 starless and 46 with associated
\iras\ sources. 
The criteria for determining whether a core has an associated
``star," meaning a central luminosity source at any evolutionary stage,
are similar to those defined by \cite{Lee 1999}.

The molecular cloud complexes and isolated cores will be observed with
both \irac\ and \mips\ at all wavelengths from 3.6 to 160 \micron,
but we expect the 160 $\mu$m detector to be saturated toward these regions.
The $3\sigma$ limits (based on pre-launch estimates) for the scan maps 
yield a limit
on \lbol\ from 3.6 to 70 \micron\ of about \eten{-3} \lsun\ at 350 pc.
The \irac\ bands would detect a 1 Myr old, 5 \mjup\ brown dwarf at 350 pc, 
based on models \cite{Burrows 1997}. 

%\subsection{Ancillary and Complementary Data}

Through a variety of collaborations, we are extending the data base to
both longer and shorter wavelengths. The large clouds and isolated
cores are being mapped in the continnum at $\lambda \sim 1$ mm, and
in spectral lines. Selected regions have been observed with millimeter-wave
interferometers.
Further complementary data on the northern clouds in
molecular lines and dust extinction and emission are being obtained by the 
\complete\footnote{http://cfa-www.harvard.edu/$\sim$ agoodman/research8.html}
team, who also plan to make the data public.
In addition, the large clouds and many northern cores are
being imaged at shorter (R, i, and z) wavelengths to limiting
magnitudes of 24.5, 22, and 22, respectively.  

\subsection{The Evolution of Disks up to 10 Myr} \label{stars}

%\subsection{Scientific Questions}

Circumstellar disks are the likely sites of planet
formation. While disks around T Tauri stars have been studied extensively, 
much less is known about both very early stages in the formation
of disks and their later evolution after accretion has ended.
The c2d program on cores and clouds can detect forming 
disks at very early stages
(disks as small as 1 M$_{Earth}$ can be detected through $\av = 100$ mag).
We will also target ``debris disks,'' which 
are found in association with nearby stars at ages up to about 1 Gyr.
We can detect such disks with very small amounts of dust ($\sim 0.1 M_{Moon}$).
The timescales for dispersal of disks in such systems
are largely unknown.

We will target weak-line T Tauri stars (wTTs) to study disk evolution
up to a few Myr.
We will address the following key scientific questions.  
What is the frequency of debris 
disks in the low-mass stellar population at ages of less than 5 Myr?  
When does the transition from primordial to debris disk occur?
Assaying the full range of disk properties at a given age will
help to distinguish evolutionary effects from other factors.
Do most disks develop inner holes or gaps during the dissipation process?
We will detect evidence for gaps using the \sed; the longer wavelengths
accessible by \sirtf\ can reveal gaps in the main planet-building zone.

We restricted our sample to wTTs stars that lie within $5\degree$ of the 
nearest large molecular clouds that we are mapping. We also required evidence of
youth in the form of X-ray emission or strong Li I 6707 \AA\
absorption. These criteria should allow us to extend the evolutionary
stages studied in the clouds to objects with ages up to about 10 Myr.
We have worked with the \feps\ team to ensure a smooth transition to
their program.
We will detect excess emisison by comparison to \sirtf\ standard stars
and model photospheres.
The survey will readily detect disks like
that around $\beta$ Pic, even if they have evacuated inner holes with 
radii as large as 30 AU.  Disks an order of magnitude more tenuous will 
still be detected if their inner edge is at a radius of 5 AU.  
We have obtained spectra of the stars in the visible to determine stellar
properties and observations with  adaptive optics to search for multiplicity.

%irs.tex  This covers the spectroscopy.

\subsection{The Evolution of the Building Blocks of Planets} \label{irs}

%\subsection{Scientific Questions}

Our spectroscopic program with the \irs\ instrument will provide 
crucial information on the evolution in the nature of the materials
that build planets, the dust, ice, and gas in circumstellar envelopes
and disks.  The mid-infrared wavelength range has a wealth of diagnostic
features for these three components. \iso\ revealed the potential
of such observations for more massive stars, and
\sirtf\ will allow spectroscopic studies of solar-type stars.
In the 75 hour c2d \irs\ program, high
signal-to-noise spectra will be obtained over the full 5--40 $\mu$m range 
[high resolution ($R\approx 600$) over the 10--37 \micron\ range]
for all phases of star- and
planet-formation up to ages of $\sim$5 Myr for at least 170 sources.
The MIPS-SED mode at 50--100 $\mu$m will also be used in the second
year of the program to characterize the longer wavelength silicate and
ice features of a disk sub-sample. 

The \irs\ spectra can be used to address the following questions. 
How do spectroscopic diagnostics evolve through the different
stages of early stellar evolution?  
How does the chemical composition of dust and ices change from
molecular clouds to planetary bodies?  
Mid-infrared spectroscopy of disks as the envelope is clearing will
be particularly interesting.
In the later evolutionary stages, changes in silicate features become
the main diagnostics.  
The spectra will form a powerful data base to compare with spectra of
Kuiper-Belt objects, comets and asteroids obtained in other \sirtf\
programs and to clarify the links between interstellar, circumstellar and
solar-system material.
How does the size distribution of dust grains evolve in
circumstellar environments?  
These observations will permit
independent estimates of the dust coagulation and gas dissipation
time scales, processes of great importance for planet formation.
What is the spectral evolution of substellar objects?  
The young brown dwarfs and super-Jupiters discovered in the
IRAC and MIPS surveys ($\sim$ 100 expected) will be targets for
follow-up observations.

%\subsection{Sample Selection}

We will reserve roughly half the \irs\ time for follow-up observations
of objects found in the mapping programs. The remaining half is scheduled
for observations of known embedded and
pre-main-sequence stars. We can study objects down to 
a typical source luminosity of $\rm 0.1~L_{\odot}$,
allowing study of masses down to $\rm 0.1~M_{\odot}$ at an age of 1 Myr.
We plan to achieve a signal to noise of 50--100 on the continuum.
This level will allow us to study the thermal history of the
envelope through the 15 $\mu$m CO$_2$ bending mode profiles 
and to search for gas phase emission and absorption
features in all phases of star formation. 
To obtain data with higher spectral resolving power in atmospheric windows, 
we are conducting flux-limited surveys in the L- and M-band windows using
the VLT-ISAAC and Keck NIRSPEC instruments.
A subset of IRS targets with known infrared ice absorption
features is being characterized, using the CSO and OVRO facilities.
The c2d-\irs\ team will enhance the \irs\ pipeline data delivered by the
SSC in several ways. The most important improvement will be in the
defringing of the IRS spectra, because laboratory experiments indeed
show the presence of fringes.

\section{Summary} \label{summary}

The c2d program will provide a legacy for future research on star and
planet formation. We hope to provide a data base for unbiased
statistical studies of the formation of stars and substellar objects.
Data from other wavelength regimes will add to the picture, and 
modeling tools will assist researchers in using the data base.
Follow-up studies of these samples with \sirtf\ itself and with future 
missions is a natural outcome.
The \sirtf\ and ancillary data will be available to the broader community
from the {\it SIRTF Science Center} (\ssc), via their Infrared Sky
Archive (IRSA).
Complementary data products will be made available, as far as possible, 
through either IRSA or public web sites.
Further information on the program, including the source lists,
can be found at the 
c2d website: {\tt http://peggysue.as.utexas.edu/SIRTF/}.

%\acknowledgements

This material is based upon work supported by the National Aeronautics
and Space Administration under Contract No. 1224608 issued
by the Jet Propulsion Laboratory.
The Leiden SIRTF legacy team is supported by a Spinoza grant from the
Netherlands Foundation for Scientific Research (NWO) and by a grant from
the Netherlands Research School for Astronomy (NOVA). 

%%%%%%%%%%%%%%%%%%%%%%%% referenc.tex %%%%%%%%%%%%%%%%%%%%%%%%%%%%%%
% sample references
% "physics"
%
% Use this file as a template for your own input.
%
%%%%%%%%%%%%%%%%%%%%%%%% Springer-Verlag %%%%%%%%%%%%%%%%%%%%%%%%%%

%
% BibTeX users please use
% \bibliographystyle{}
% \bibliography{}
%
% Non-BibTeX users please use

%%%%%%%%%%%%%%%%%%%%%%%%%%%%%%%%%%%%%%%%%%%%%%%%%%%%%%%%%%%%%%%%%%%%%%  }

%%%%%%%%%%%%%%%%%%%%%%%%%%%%%%%%%%%%%%%%%%%%%%%%%%%%%%%%%%%%%%%%%%%%%%

\printindex
\end{document}